\documentclass{elsart1p}

\usepackage{amssymb}
\usepackage{epsfig}

\begin{document}

\begin{frontmatter}



\title{Renormalisation of out-of-equilibrium quantum fields}


\author{Sz.~Bors\'anyi $^a$ and U.~Reinosa $^b$}

\address{$^a$ Department of Physics and Astronomy, University of Sussex,\\
Brighton, East Sussex BN1 9QH, United Kingdom;\\
$^b$ Centre de Physique Th{\'e}orique, Ecole Polytechnique, CNRS,\\
91198, Palaiseau, France.}

\begin{abstract}
We consider the initial value problem and its renormalisation in the framework of the two-particle-irreducible (2PI) effective action. We argue that in the case of appropriately chosen self-consistent initial conditions, the counterterms needed to renormalise the system in equilibrium are also sufficient to renormalise its time evolution. In this way we improve on Gaussian initial conditions which have the disadvantage of generically not showing a continuum limit. For a more detailed discussion see \cite{Borsanyi:2008ar}. 
\end{abstract}

\end{frontmatter}

The two-particle-irreducible (2PI) effective action \cite{Baym} provides an efficient first principles approach to out-of-equilibrium quantum field theory which circumvents the secularity problem and allows to study late-time-dynamics in scalar theories \cite{CoxBerges} as well as in models incorporating fermions \cite{FermionThermalisation}. It has by now become a standard framework for nonequilibrium quantum field theory with mostly scalar applications of
cosmological interest \cite{HeidelbergParametric}. An open question remains however, namely how to initialize the system in such a way that a continuum limit be defined.

\begin{figure}[htpb]
\centerline{\includegraphics[width=1.7in]{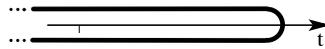}}
\caption{Real-time path.\label{fig:rtp}}
\end{figure}

For illustration, let us consider a scalar $(\lambda/4!)\varphi^4$ theory. In equilibrium, the system can be described on the so-called real-time path, see Fig.~\ref{fig:rtp}, on which
the two-point function $G$ can be decomposed in terms of the statistical propagator $F$ and the spectral density $\rho$. These fulfill the equations of motion
\begin{eqnarray}
\left(\partial_x^2+m^2+\Sigma_{0}\right)F(x) & = & \int\limits^{0}_{-\infty} \!dz^4\,\Sigma_F(x-z)\rho(z)
-\int\limits^{x_0}_{-\infty} \!dz^4\,\Sigma_\rho(x-z)F(z)\,,\label{eq:c}\\
\left(\partial_x^2+m^2+\Sigma_{0}\right)\rho(x) & = &
- \int\limits^{x_0}_0 dz^4\,\Sigma_\rho(x-z)\rho(z)\,,\label{eq:d}
\end{eqnarray}
where, in the 2PI approach, the self-energy $\Sigma$ is obtained from the sum $\Gamma_{\rm int}[\phi,G]$ of two-particle-irreducible diagrams as $\Sigma=2i\delta\Gamma_{\rm int}/\delta G$ \cite{Baym,Borsanyi:2008ar}. In order to yield a particular solution, these equations need to be supplemented by appropriate boundary conditions: for the spectral density these are fixed by the equal time commutation relations; for the statistical propagator one should use the KMS condition (in Fourier space)
\begin{equation}\label{eq:KMS}
F(p_0;\vec{p})=-i\left(\frac{1}{2}+\frac{1}{e^{\beta p_0}-1}\right)\rho(p_0;\vec{p})\,,
\end{equation}
which sets the temperature $\smash{T=1/\beta}$ of the system in equilibrium. The KMS condition also ensures that the approach on the real-time path is equivalent to approaches on contours with decreasing imaginary part and stretching from a time $t_I$ to a time $t_I-i\beta$, more directly connected to the canonical or path integral formulations of QFT.

If the KMS condition is assumed, Eqs.~(\ref{eq:c}) and (\ref{eq:d}) become strictly equivalent and it is enough to solve for $\rho$ in Eq.~(\ref{eq:d}) and determine $F$ from Eq.~(\ref{eq:KMS}). Still, if one aims at defining an equilibrated state in the ``continuum limit'', Eq.~(\ref{eq:d}) needs to be renormalised.\footnote{We assume the Landau pole to be much higher than the relevant scales in the problem. By ``continuum limit'' we mean that the regularisation cut-off is taken to much larger values than these relevant scales but much lower values than the Landau pole.} Renormalisation in the 2PI approach for a system in equilibrium is well known by now in theories including scalar \cite{HeesKnoll}, fermionic \cite{ReinsaFermions} as well as gauge \cite{RSQEDRenorm} degrees of freedom and amounts to introducing appropriate counterterms in the equations of motion together with appropriate renormalisation conditions. In moving to out-of-equilibrium situations, one should keep in mind that these counterterms, which are time-independent, should also be present in the evolution equations, if one wants the system to evolve toward the correct ``continuum'' equilibrated state. Then, as we illustrate shortly, depending on how one puts the system into motion, these counterterms can yield unbalanced divergences which might spoil the dynamics.\\

A possible way to bring the system out-of-equilibrium in the real-time formalism is to introduce a source term $K_F$ in the equations of motion
\begin{eqnarray}
\left(\partial_x^2+m^2+\Sigma_{0}(x)\right)F(x,y) & = &
\int\limits^{y_0}_{-\infty} \!dz^4\,\Sigma_F^K(x,z)\rho(z,y)-\int\limits^{x_0}_{-\infty} \!dz^4\,\Sigma_\rho(x,z)F(z,y)\,,\label{eq:a}\\
\left(\partial_x^2+m^2+\Sigma_{0}(x)\right)\rho(x,y) & = &-\int\limits^{x_0}_{y_0} dz^4\,\Sigma_\rho(x,z)\rho(z,y)\label{eq:b}\,,
\end{eqnarray}
where $\smash{\Sigma_F^K\equiv\Sigma_F+iK_F}$ and
\begin{equation}
K_F(x,y)\equiv\left\{\begin{array}{ll}
K_F(x-y)&\textrm{if $x_0<0\,$ and $\,y_0<0$\,;}\\
0&\textrm{if $x_0>0\,\,\,$ or $\,\,\,y_0>0$\,.}
\end{array}\right.
\end{equation} 
For times smaller than zero, the system is in a steady state sustained by the
translationally invariant source $K_F(x-y)$. Such states exist for all times
and can be obtained by solving the simplified, translationally invariant
equations
\begin{eqnarray}
\left(\partial_x^2+m^2+\Sigma_{0}\right)F(x) & = & \int^{0}_{-\infty} \!dz^4\,\Sigma_F^K(x-z)\rho(z)-\int^{x_0}_{-\infty} \!dz^4\,\Sigma_\rho(x-z)F(z)\,,\\
\left(\partial_x^2+m^2+\Sigma_{0}\right)\rho(x) & = & -\int^{x_0}_{0} dz^4\,\Sigma_\rho(x-z)\rho(z)\,.
\end{eqnarray}
The source $K_F(x-y)$ alone does not completely fix the solution of these equations, as it was also the case
in equilibrium (no source). To define a particular solution we
introduce a non-thermal boundary condition in the form of a generalized
KMS condition
\begin{equation}\label{eq:KMS_gen}
F(p_0;\vec{p})=-i\left(\frac{1}{2}+f(p_0;\vec{p})\right)\rho(p_0;\vec{p})\,,
\end{equation}
where $f(p_0;\vec{p})$ is an arbitrary function. After such a steady state has been obtained, one can evolve the system for times greater than zero by solving Eqs.~(\ref{eq:a}) and (\ref{eq:b}) where $K_F$ has been set to zero and where in the integrals, the contributions from times smaller than zero involve the pre-calculated steady state. 

This prescription, which we name self-consistent for the corresponding source $K_F$ depends on the two-point function $G$, see Eq.~(\ref{eq:source}) below, has to be compared to the usual 2PI evolution equations solved using Gaussian initial conditions which amount in particular to start the dynamics at a time $\smash{t=0}$ on the real-time path, with the memory integrals in Eq.~(\ref{eq:a}) starting at $\smash{t=0}$ rather than at $\smash{t=-\infty}$. In this latter case, the counterterms which are time-independent and thus present at initial time, produce an unbalanced initial time singularity, which aside from very specific examples, can hardly be removed. Our approach in contrast, for it involves memory integrals running from $\smash{t=-\infty}$ produces contributions which could compensate the counterterms at any time. For this to be true however the source $K_F(x,y)$ needs to have a safe UV behavior in order not to bring additional divergences as compared to the situation without source.\footnote{For a proof of this statement in Euclidean time, see Ref.~\cite{Borsanyi:2008ar}} For example, in the definition of the steady state, the generalized KMS condition (\ref{eq:KMS_gen}) corresponds to a source $K_F(x-y)$ such that \cite{Borsanyi:2008ar}
\begin{equation}\label{eq:source}
K_F(p_0;\vec{p})=-\left(\frac{1}{2}+f(p_0;\vec{p})\right)\Sigma_\rho(p_0;\vec{p})
+i\,\Sigma_F(p_0;\vec{p})\,.
\end{equation}
Thus by choosing for instance
\begin{equation}
f(p_0;\vec p)=\frac1{e^{p_0/T(\vec p)}-1}
\end{equation}
where $\smash{T(\vec{p})}$ converges fast enough in the UV to a reference temperature $T^\star$, we ensure that the UV properties are not modified with respect to equilibrium at temperature $T^\star$ and thus that the equations of motion for the steady state are renormalised by the very same counterterms as in equilibrium.

\begin{figure}[htbp]
\includegraphics[width=2.6in]{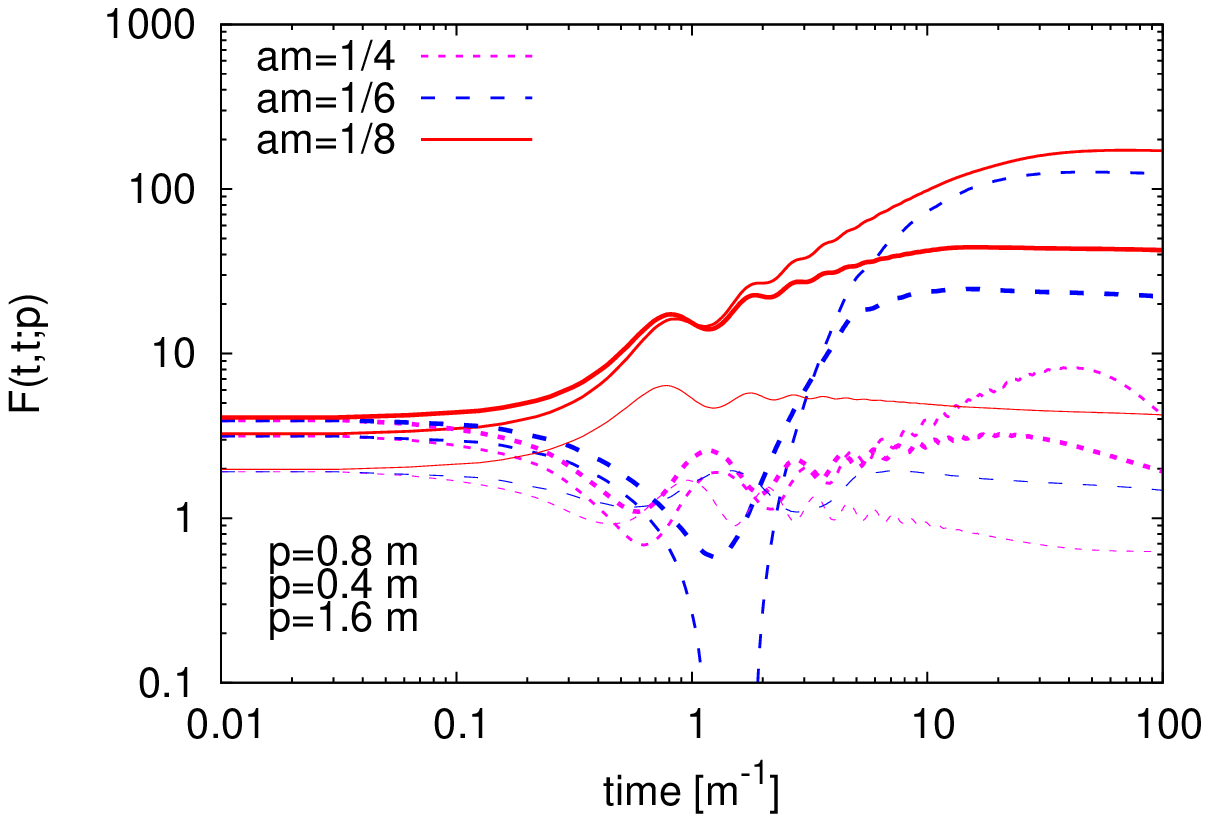}$\;$\includegraphics[width=2.6in]{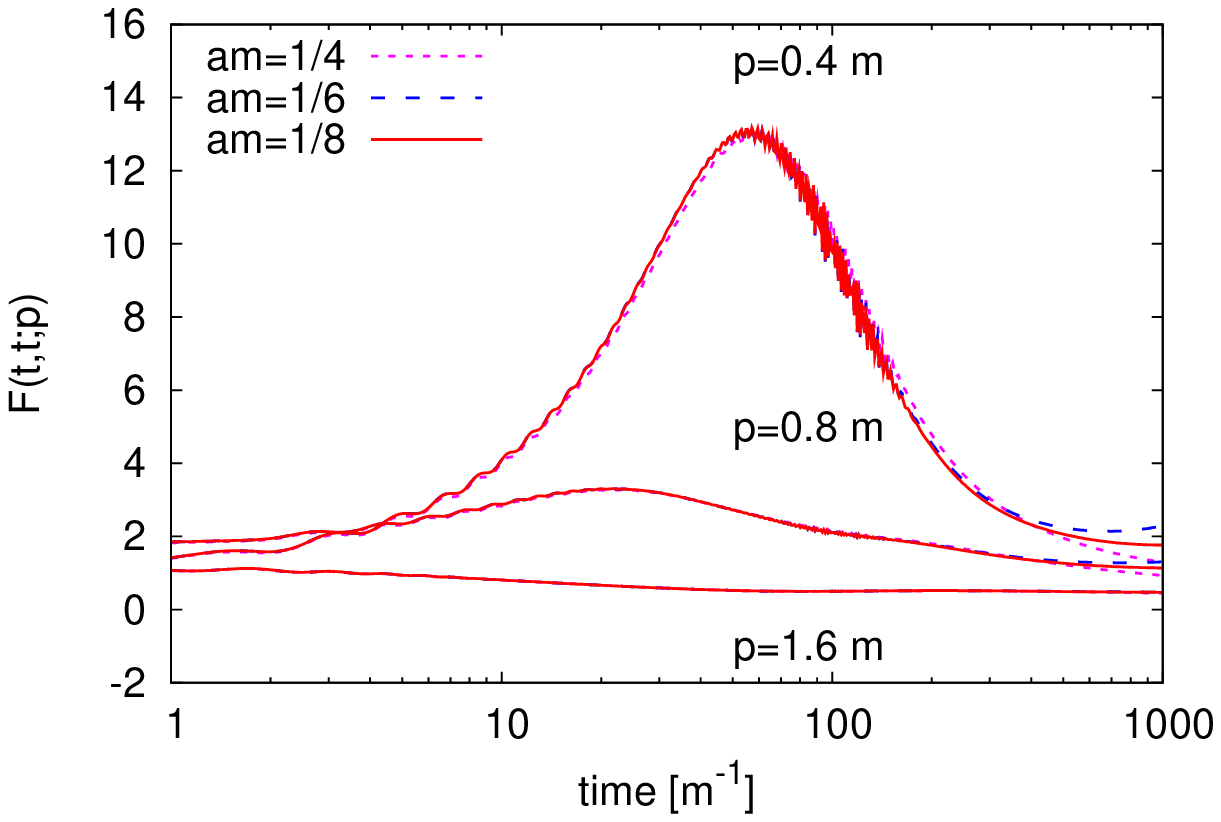}
\caption{
Gaussian versus self-consistent initial conditions ($\smash{\lambda=24}$, $\smash{\mbox{box size } 32}$). See the text for details.\label{fig:contlim}
}
\end{figure}

In Fig.~\ref{fig:contlim} we compare the time evolution of the equal-time
statistical propagator $F(t,t;\vec p)$ choosing either Gaussian or self-consistent initial conditions. We consider three different modes $\smash{|\vec{p}|=0.4,0.8,1.6}$ and three different values of lattice spacing $am=1/4,1/6,1/8$.  The counterterms are determined once and for all in equilibrium at the reference temperature $\smash{T^\star=m}$. The absence of a continuum limit is apparent in the case of Gaussian initial conditions. As for self-consistent initial conditions, for a given mode, the curves representing runs for different values of the regularisation cut-off lie almost exactly on top of each other, strongly suggesting that the continuum limit has been reached.


\vspace{-0.5cm}


\end{document}